\documentclass[twocolumn,showpacs,prl]{revtex4}
\usepackage{dcolumn}
\usepackage{bm}

\begin{document}

\title{On the Possibility of Large Upconversions and Mode Coupling between
Fr\"{o}hlich States and Visible Photons in Biological Systems}

\author{J. Swain}

\affiliation{Physics Department, Northeastern University \\
110 Forsyth Street, \\
Boston, MA 02115, USA \\
and \\
International Institute of Biophysics (IIB)\\
Landesstiftung Hombroich, Raketenstation\\
D-41472 Neuss, Germany\\
E-mail: john.swain@cern.ch}

\begin{abstract}
At least two significant roles for large scale quantum coherence in
living systems have been suggested: Herbert  Fr\"{o}hlich's 
coherent excitations of
nonlinearly coupled ensembles of large polarizable molecules,
with frequencies in the microwave region, and Fritz Popp's
coherent visible photon
emission arising from metabolically active cells.
The large difference in frequencies has made it difficult to
see these two phenomena as being linked. Here a case is made
for a potentially large coupling between these modes,
and suggestions are made for the biological roles played by these coherent
excitations.
\end{abstract}

\maketitle

\section{Introduction \label{intro}}

This paper is based on contributions I made to a joint paper\cite{jointpaper} to be
presented by Fritz Popp at the Fr\"{o}hlich Centenary Symposium
in Liverpool in April 2006.

There are two main streams of thought which suggest macroscopic quantum mechanical effects
associated with the living state. One is Herbert Fr\"{o}hlich's
suggestion\cite{Frohlich,Frohlich-selective} that one would expect
coherent
excitations in the microwave region of the spectrum due to nonlinear couplings of biomolecular
dipoles -- a suggestion which is perhaps more theoretically developed than experimentally
investigated. For brevity, I will refer to microwave photons in such a state as 
``Fr\"{o}hlich photons''. The other is Fritz
Popp's observation\cite{Popp} that the photocount statistics of ultraweak bioluminescence (biophotons) around the
visible region of the spectrum, and other data,  suggest a coherent component linked to the living
state. Here it is important to note that one is not forgetting that there are plenty of reactions -- in
particular oxidative ones  -- which will give rise to incoherent photoemission and are simply a part of
the normal chemistry that goes on in living things. Such a traditional contribution would always be
expected, but here we are considering the rather remarkable, spectrally flat distribution which seems to
be associated with life. I will refer to visible photons of this kind as ``Popp photons''.

Fr\"{o}hlich's original model was based on phenomenological rate equations representing a set of harmonic
oscillators labelled by an index $i$ of various frequencies, $\omega_i$ ,
each excited with $n_i$ quanta, and coupled to each
other. In order to be able to couple these oscillators and still have conservation of energy, the
oscillators are also coupled to a heat bath at some temperature T which provides or accepts whatever
energy is needed ({\em i.e.} $\hbar(\omega_1-\omega_2)$ for couplings
of oscillators of frequencies $\omega_1$ and $\omega_2$). In addition, an external pump source was included which could drive the oscillators.

Fr\"{o}hlich showed that under rather general conditions one could, even at high temperatures, for a
sufficiently strong external pump, find a macroscopic excitation of just one oscillator -- the one with
the lowest energy  -- at the expense of the other modes. It was this macroscopic quantum state, which is
reminiscent of lasing, or of Bose-Einstein condensation, which has been identified as possibly
characteristic of the living state. While Fr\"{o}hlich came upon this sort of mechanism while thinking about
the very large electric fields across the membranes of living cells, it is important to keep in mind
that this could occur multiple times in different systems in one living organism, and one should be open
to the idea of multiple Fr\"{o}hlich states appearing in various contexts. That being said, if one imagines
the dipoles in question to be large biological molecules, one is naturally led to frequencies in the
microwave region and it is far from clear what, if anything, this has to do with biophotons.

Before dismissing any connection right away due to the fact that there is a difference in energy scales
between Fr\"{o}hlich and Popp photons, let us recall that large upconversions are, in fact, well-known,
although rarely considered in any detail in the standard texts. For example, consider a light bulb running from a 1.5 V
battery. There are two natural energy scales going into the system. One is 1.5 eV -- the total energy that
an electron can give up as it crosses the filament of the bulb, and the other is zero (!) -- the
frequency of the input current (which is DC here).  Even the 1.5 eV is rather optimistic since one
expects an electron to undergo many collisions as it passes through the filament, each time giving up
only a small fraction of an eV. Nevertheless, common experience shows that there is strong emission in
the visible spectrum of several electron volts. That is, a macroscopic nonlinear system, even without
coherence, manages to couple many low energy quanta into one higher energy quantum of, say, blue light.
A first response on hearing this argument is to trivialize it, saying ``Well, isn't that just blackbody
radiation?''. The answer, of course, is yes, but whenever one treats the approach to blackbody radiation
one has to imagine complicated couplings between many oscillators which one subsumes into statements
like ``one lets the oscillators exchange energy until the most probable distribution (via maximizing
entropy) is reached''. With this in mind, it may come as less of  a surprise that a highly organized
system may be more efficient at large upconversions.

Let us now slightly generalize Fr\"{o}hlich's model (J. Swain, August 2005, talk at the Fr\"{o}hlich
Symposium, IIB, Neuss, Germany)\cite{Swain2005} and add one more term to his rate equations. This will be a coupling of the whole
Fr\"{o}hlich system (which is a large number of oscillators all coupled together, with the lowest energy
oscillator coherently excited) to one external oscillator.  The system is already coupled
to an entire heat bath of oscillators, so this is just one more oscillator. The difference is that the
entire heatbath of oscillators in Fr\"{o}hlich's model are {\em constrained} to have their energy thermally
distributed at a temperature T. Here we add just one oscillator, not requiring it to be a part of a
thermal ensemble. Thinking in concrete terms, this can be an oscillator corresponding to the QED vacuum
and its occupation number represents the number of photons in that state. It could also be an
oscillator  corresponding to the excitation energy for some chemical reaction, or some other process.
Such an additional coupling outside the original Fr\"{o}hlich model will drain energy  from the system,
which we can compensate for by an increase in pumping so as to not lose the coherent Fr\"{o}hlich
excitations.

Now let us consider the probability that $N$ Fr\"{o}hlich photons of frequency $\omega_F$ can couple to a Popp photon
of frequency  $\omega_P$ ( equal to $N\omega_F$ ). Here is it useful to think of Feynman diagrams
which represent quantum
mechanical amplitudes for such photons to couple. Think then of a line which represents a charged
particle in a living system. This can be an electron, or a dressed quasiparticle. In keeping the
Fr\"{o}hlich's original work which stressed the generality of a mechanism rather than details of a specific
model, we will refer to this as an ``electron'', keeping in mind that generalizations are possible
(a dipole, a quasiparticle, {\em etc.}). Diagrams
contributing to the process we want have $N$ incoming photon lines and one outgoing photon line. The
amplitude for the process has a factor of electron charge (e) for each photon, and the transition
probability is obtained by adding all diagrams with all permutations of incoming photons, finding the
squared magnitude of the resulting complex quantity, and multiplying by the amount of phase space
available for the outgoing photons. In general, this is a difficult problem to work out, and one expects
(assuming that different diagrams have different phases and are unlikely to constructively interfere) a
result of order  $\alpha^n$ where  $\alpha$ is the fine structure constant -- a pure number roughly equal to 1/137.
It would then seem that
coupling of a large number $n$ of photons will be very unlikely, and it is this sort of intuition which
tends to make one immediately dismiss large upconversions as impossible or unlikely.

The Fr\"{o}hlich state, however, is a special one in that it is a coherent state in which all the photons
carry the same energy. In fact, as a coherent state, it does not have a well-defined number of photons at
all, but for the following we can consider projecting out a certain number $N$ by the requirement that
$N$ must be combined in order to get the single photon of frequency   (details will appear elsewhere, Swain,
2006, in preparation, but they are not difficult). Now we have $N$ indistinguishable photons coming in and
N! diagrams all of which must be counted and all of which have the same numerical value and thus which
interfere constructively. This is nothing more than Bose-symmetrization (with an attendant $1/sqrt{N!}$
for $N$ identical incoming photons -- that is, one must add each diagram corresponding to an indistinguishable process
(corresponding to different photon line orderings) with the same amplitude. Now the original 
naive estimate of a rate proportional to  must be replaced by $\alpha^N N!$ . For small $N$,
the $N!$ term is negligible compared to the rapid falloff in powers of the fine structure constant, but
for $N$ of about 400,  $\alpha^N N!$ becomes unity and the factorial term begins  to dominate and in fact the
perturbative concepts become unreliable. As $N$ goes to infinity, the probability of this process grows
without bound, and violates unitarity, signaling that one has pushed perturbation theory beyond its
limits of validity. This sort of breakdown of perturbation theory is in fact to be expected, as it is well-known that
the perturbative QED expansion is divergent\cite{Dyson} and in fact it is still quite mysterious that the results of
low order perturbation theory are as good as they are! It should be noted here that one cannot extend
the argument to suggest the the outgoing photon energies would grow without limit. With perturbation
theory breaking down one would need additional non-perturbative effects to set in and enforce unitarity.
Qualitatively, however, one might expect a low probability for a few photons to couple, a large
probability for about 400, and a flattening off thereafter with increasing numbers. 

In all honesty one must be aware that one is really using the structure of perturbation theory as a
guide rather than going a full non-perturbative calculation. Interesting though this is,  a
non-perturbative treatment is in preparation\cite{Swain2006} which suggests a similar result from a
different point of view.

A similar factor of $N!$ was anticipated on phenomenological grounds by Popp {\em et al.}
\cite{Slawinski}, but here finds its expression rooted in a microscopic model
linking Fr\"{o}hlich and Popp photons. Rather remarkably, the factor of 400 or so derived above is about
what one would want to get from Fr\"{o}hlich frequencies in the microwave region to visible Popp photons! The
picture which emerges then, subject to more (nonperturbative) theoretical work and, of course,
experimental verification, is:

\begin{itemize}
\item there are at least two types of quantum electromagnetic systems associated with the living state.
One is in the microwave frequency range as suggested by Fr\"{o}hlich, and one is around the visible region as
suggested by Popp
\item there is a coupling of these two which is natural within the context of known physics, assuming that
the Fr\"{o}hlich mechanism takes place
\item the model here is predictive in that it suggests coherent excitations in the microwave region followed
by small excitations at integer multiples of the Fr\"{o}hlich frequency, decreasing rapidly
in intensity at first
and then rising again at about 400 times the Fr\"{o}hlich frequency, then levelling off, and, eventually falling in
order to preserve unitarity. Exact details here are beyond perturbation theory and require different
mathematical methods. The physics here is, however, experimentally accessible, at least over part of the
frequency range, with the THz region still being quite challenging in terms of instrumentation.
\end{itemize}

There is even a degree of reversibility in the system and one can imagine coupling what we have so-far
considered as an outgoing photon mode to an external visible photon which could in turn be
down-converted into Fr\"{o}hlich photons. This offers the possibility of affecting Fr\"{o}hlich dynamics  inside
a cell via visible photons injected from outside, as well as to the possibility of a long range coupling
of internal cellular Fr\"{o}hlich dynamics between cells via visible Popp photons.This also suggests new
experiments looking for microwave responses to visible stimulation and vice-versa.

One might well ask what use such a setup with two such disparate frequency ranges is in living things.
An immediate suggestion is that living things require a large range of quanta of energy to drive various
reactions. This is akin to going shopping where one needs to make purchases using a wide range of sums
of money. In order to ensure the correct energy for each reaction (correct change, in the monetary
analogy), one would do well to store one's energy (money) in small quanta (small coins). There is a
smallest sensible quantum to store (smallest coin to keep) set by what quantity can be had or lost for
essentially nothing. This is about kT of energy for a biological system which is bathed in quanta of this
energy ({\em i.e.} is in a heat bath of temperature T) and a penny in the money analogy since nobody really cares about getting or losing one. Storing
energy in microwaves ensures that one can be very close to just about any amount of energy that would be
needed to drive an electronic transition (chemical reaction) as needed. Storing in lower energy
excitations makes no sense then, and higher frequencies would, perhaps surprisingly, not be expected to
couple as easily to the required energy scale. Even a degree of automatic specificity is expected since
a reaction which is all set to go but for the energy required constitutes a resonance (extra phase
space) into which the outgoing quantum can couple. This is complementary to Fr\"{o}hlich's 
``selective long range dispersion forces between large systems''\cite{Frohlich-selective} now extended beyond physical
forces to selective transfer of energy. In other words, there is a natural framework not just for
a biological molecule to experience long-range forces pulling it to where it should go, but also
for the appropriate amount of energy to be transferred between them.
The use of single photons as
part of cell-to-cell signalling is also fascinating and the sort of system here could allow for a high
degree of selectivity with little cross-talk by choosing slightly different optical frequencies for
different communications. Of course just what nature does is an experimental question, and likely to
provide even more surprises!

\section{Acknowledgements}

I would like to thank the NSF for continuing support of other research,
and the members of the IIB in Neuss, and especially Fritz Popp,
for fascinating discussions.

\end{document}